\documentclass{taira}

\usepackage{amssymb}
\usepackage{amsmath}
\usepackage{graphicx}
\usepackage{lipsum}

\begin{document}

\shorttitle{Active flow control of wall-normal vortex} 
\shortauthor{Q. Liu et al.} 

\title{Active flow control of a pump-induced wall-normal vortex with steady blowing} 

\author
 {
  Qiong Liu\aff{1}
  \corresp{\email{qliu3@ucla.edu}},
  Byungjin An    \aff{2},
  Motohiko Nohmi \aff{2},
  Masashi Obuchi \aff{3},
  \and
  Kunihiko Taira \aff{1}
  }

\affiliation
{
\aff{1}
Department of Mechanical and Aerospace Engineering, University of California, Los Angeles, CA 90095, USA

\aff{2}
Fundamental Technologies, Research Department, Ebara Corporation, Tokyo,144-8510, Japan

\aff{3}
Advanced Analysis Department, Ebara Corporation, Tokyo, 144-8510, Japan }

\maketitle

\begin{abstract}

The emergence of a submerged vortex upstream of a pump can reduce pump intake efficiency and cause structural damage.  In this study, we consider the use of active flow control with steady blowing to increase the pressure distribution within a single-phase pump-induced wall-normal vortex model, which is based on the Burgers vortex with a no-slip boundary condition prescribed along its symmetry plane.  The goal of our control is to modify the vortex core velocity profile.  These changes are sought to increase the core pressure such that detrimental effects on the pump are alleviated. Three-dimensional direct numerical simulations (DNS) are performed to examine the dynamics of the vortex with the application of axial momentum injection at and around the root of the vortex. We find that the active flow control approach can effectively modify the wall-normal vortical structure and significantly increase the low-core pressure by up to $81\%$ compared to that of the uncontrolled case. The result shows that the control setup is also effective when it is introduced in an off-centered manner. Compared to the unsteady blowing and suction based actuation from our previous work \citep{liu2018core}, the current steady control technique offers an effective and simple flow control setup that can support robust operations of pumps.
\end{abstract}

\section{Introduction}
\label{sec:intro}

The requirement of transporting water has become ever more challenging with the increased occurrence of heavy storms and floods.  Since the vast majority of the human population is inhabited near the coast and river lines, it is absolutely critical that engineering infrastructures do not fail to control water levels.  One of the critical components in the effort to control floods and surging water is the pump sump.  The pump sump settles the water to be removed from a system prior to its transport by a pump.  While the pump sump and the pump are designed for a range of conditions, off-design operations call for a careful examination of the pump sump and the pump to ensure stable and efficient operations.

In off-designed conditions, strong vortices can appear in the pump sump.  For flood water control, the incoming water often contains debris and air bubbles, which further challenges the pumping operation.  In these cases, the strong vortices that emerge in the pump sump can possess hollow vortex cores.  These vortices can appear submerged in the water or connected to the free surface, whose presence can significantly degrade the pump intake efficiency and potentially cause structural damage from unbalanced loads \citep{Posey1950, Zhao:Vis10, Brennen11, Yamada, nagahara2001effect, An:Ebara18}. 

In an effort to weaken and suppress vortices in pump sumps, there have been numerous studies on the implementation of passive flow control devices.   For example, anti-vortex baffles have been considered\citep{yang2017computational, kim2012study}.  In particular, Kim {\it et al.}~\citep{kim2012study} studied the effectiveness of an anti-vortex device in the sump model by experiments and numerical simulations.  Their results showed that the passive anti-vortex device can enhance the intake efficiency by up to $91\%$ with reduced flow unsteadiness into the pump.  Their study also identified a challenge with passive control.  The control technique cannot easily adapt to changes in the incoming flow rate which leaves an open question on the robustness and adaptability of passive flow control for pump sump.

\begin{figure*}
\centering
\includegraphics[width=0.9\textwidth]{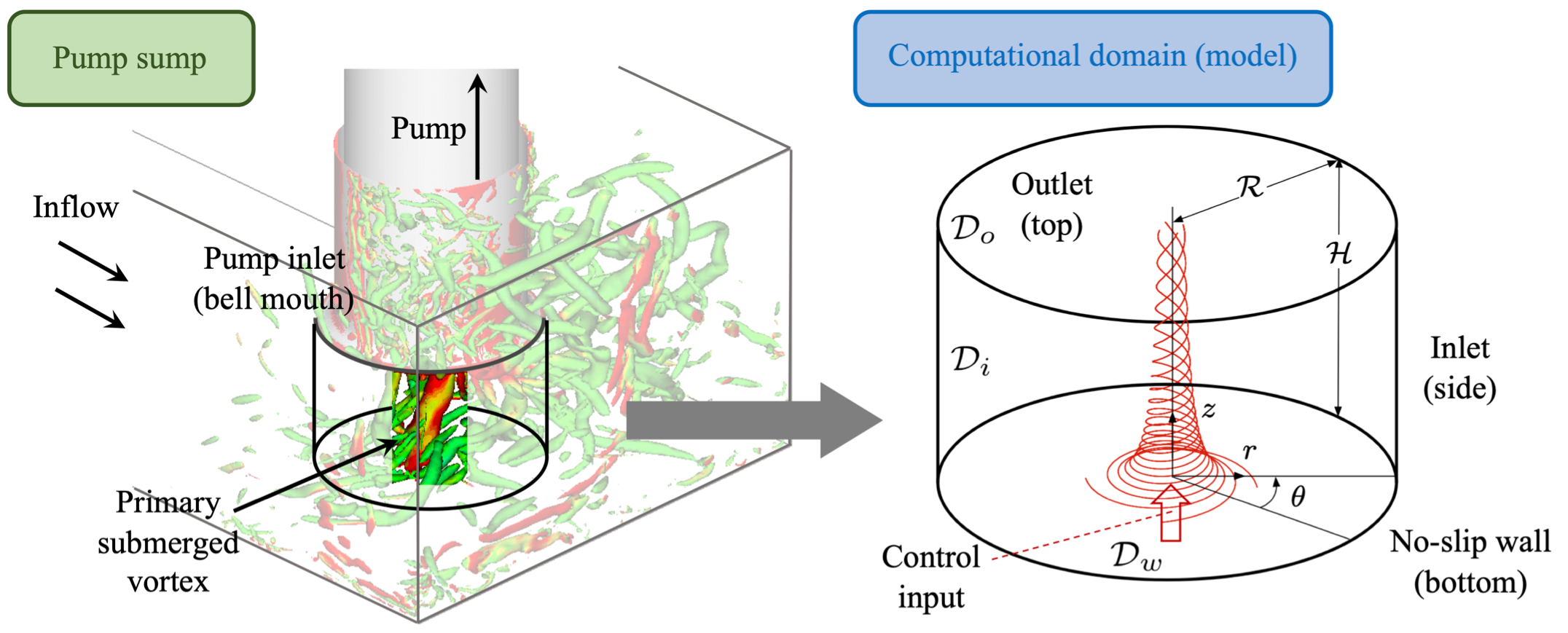}
\caption{Vortical flow in a pump sump (left) and the model flow considered in the present study (right).  Isosurface of $Q$ and transparency used for the left visualization to highlight the primary vortex.}
\label{fig:geometry} 
\end{figure*}

More recently, active flow control has been considered for its ability to address the issues associated with passive flow control.  To highlight the challenges with controlling the flow, let us show the complexity of the vortical flow in a pump sump.  Visualized in Fig. \ref{fig:geometry} (left) is a vortical flow field obtained from a large eddy simulation with the WALE model \citep{An:APS2019}. The flow field for the shown case has a primary submerged vortex upstream of the pump inlet.  To alleviate the influence of this strong wall-normal vortex, we herein consider the application of active flow control that can adaptively introduce perturbations to modify the vortex core profile. 

%%% FIGURE 1 %%%

Previous work on vortices provides theoretical guidance on developing flow control strategies\citep{zhang2009characteristics, pasche2017part, pasche2019optimal}.  Triggering and suppression of vortex breakdown have been considered by Brown {\it et al.}~\citep{brown1990axisymmetric} to modify the base vortex.  Moreover, Shtern {\it et al.}~\citep{shtern1997vortex} developed a new class of analytical solutions of the $Q-$vortex and suggested ways to predict and control the vortical flow by adding an axial input. Their study reveals that the incorporation of the axial flow can be leveraged in modifying the core profile. Bosioc {\it et al.}~\citep{bosioc2012unsteady} and {\v{S}}tefan {\it et al.}~\citep{vstefan2017proper} experimentally studied the effect of an axial water jet injection on the vortex core pressure. Additionally, Mullin {\it et al.}~\citep{mullin2000creation} used a small rotary rod placed along the vortex core to study the behavior of the vortex over different rod rotation speeds.  Husain {\it et al.}~\citep{husain2003control} used a similar experimental setup to examine the vortex behavior by adding a small rotating rod near the axis. The controlled results showed that the co-rotating control retained the steady flow while the counter-rotating control made the flow unsteady and stimulated traveling vortex rings.

These past studies suggest that active injection of perturbations to the flow can alter the core velocity profile. Our recent work \citep{liu2018core} considered a fluid-based actuation guided by data-based modal analysis to implement a rotary synthetic jet actuation at the vortex root.  The controlled results revealed that co-rotating actuation achieves a significant increase in vortex core pressure.  With the capability to turn on actuation only when needed in a tunable manner, unsteady active mass injection is an attractive control technique.  Such a flow control scheme is of particular importance because it does not block the flow and suffer from the loss of efficiency.

Based on our previous active flow control efforts, we choose active mass injection as a promising candidate.  We aim to simplify the control scheme further using the steady mass injection, as well as assessing its robustness.  As a model problem to assess the effectiveness of the active flow control effort, we consider a wall-normal vortex based on a viscous Burgers vortex \citep{Burgers:AAM48} with a no-slip wall prescribed perpendicular to its vortex axis. The numerical setup is illustrated in Fig. 1 (right).  In what follows, we describe the baseline vortex computational setup in section \ref{sec:Problem_setup}, and present baseline vortex characteristics in section \ref{sec:Baseline}. In section \ref{sec:ctrl_setup}, we provide details on the steady control approach used for modifying the vortical flow. The controlled flow characteristics and control effect are documented in section \ref{sec:ctrl_results} using a comparison with the unsteady control counterpart. At last, concluding remarks are offered in section \ref{sec:conclusion}.

%\lipsum[2-3]

\section{Computational set-up}
\label{sec:Problem_setup}

We consider a single-phase wall-normal vortex as a model of the vortical flow that develops upstream of pump inlet. The computational setup of this flow is illustrated in Fig.~\ref{fig:geometry} (right). The flow enters the computational domain from the circumferential side ($\mathcal{D}_i$) and leaves the domain from the top outlet ($\mathcal{D}_o$), where a convective outflow condition is prescribed.  Along the bottom wall ($\mathcal{D}_w$), a no-slip wall boundary condition is applied.  For the inflow condition, a Burgers vortex velocity profile \citep{Burgers:AAM48, liu2018core} is specified with
\begin{equation}
   u_r=-\frac{1}{2}\gamma r, \quad 
   u_\theta=\frac{\Gamma_\infty}{2\pi r}\left[1-\exp\left(-\frac{r^2}{a^2}\right)\right], \quad 
   u_z=\gamma z,
\end{equation}
where $\gamma$ is the strain rate of the swirl flow and $a$ is the vortex core radius.  
In the present setup, the maximum swirl velocity reaches $u_{\theta,\max}=1$ at $r = 1.12a$.
We non-dimensionalize the variables here using $a$ for length, $u_{\theta,\max}$ for velocity, and $p^\ast=\frac{1}{2}\rho u_{\theta,\max}^2$ for pressure, where $\rho$ is the density of the fluid.
In the present study, we define a circulation-based Reynolds number $Re=\Gamma_\infty/\nu =5000$, where $\nu$ is the kinematic viscosity.  
The strain rate $\gamma$ is determined from the choice of Reynolds number and the core radius $a$ such that $\gamma = 4\nu/a^2$.  The circulation based on the present scaling is $\Gamma_\infty a/u_{\theta,\max} = 9.848$.

\begin{figure*}
\centering
\includegraphics[width=0.7\textwidth]{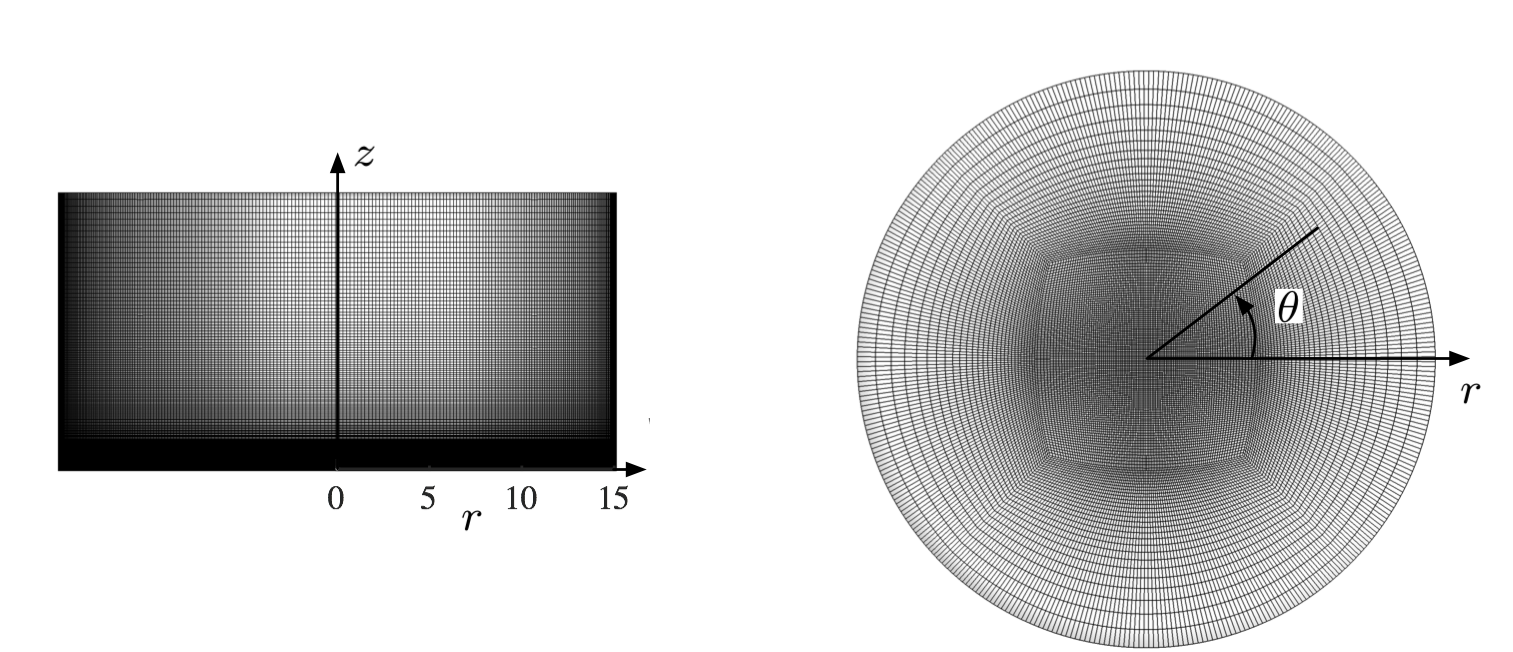}
\caption{Side and top views of the computational grid.}
\label{fig:mesh} 
\end{figure*}

Direct numerical simulations (DNS) are performed with an incompressible flow solver \emph{Cliff} (CharLES software package, Cascade Technologies) based on a collocated node-based second-order finite-volume method and a fractional-step scheme. Spatial derivatives for the fluid velocity are approximated using central differencing\citep{ham2004energy,ham2006accurate}. The computational domain is discretized into a structured mesh with finer grids clustered near the $z$-axis for resolving the vortical structures and boundary layer along the wall, as shown in Fig.~\ref{fig:mesh}.
The computational setup has been justified by examining three different computational domain sizes and grid resolutions, as summarized in table \ref{table:verification}. 
The comparisons of the minimal value of vortex core pressure, $\min (p_\text{avg}/p^\ast$), and its axial location, $z/a$, in the flow field among three cases exhibit agreement. Based on the mesh and domain size test, we choose D2 as the computational setup for the simulations.
With the appropriate choice of mesh and domain size, we now perform simulations of the baseline flow, using active flow control to modify the core structure with the aim of increasing the pressure profile.

%%% FIGURE 2 %%

%%% TABLE 1 %%%
\begin{table}
\begin{center}
\caption{Mesh and domain dependency test using the minimal time-averaged core pressure  min($p_\text{avg}/p^\ast$) and its axial location $z/a$}
\scalebox{0.85}{	\begin{tabular}{ccccc}\hline
      Cases & Domain size ($\mathcal{R}\times \mathcal{H}$) & Mesh size & min($p_\text{avg}/p^\ast$) & $z/a$ \\ \hline
      D1 & $15 \times 20$ & 22,657,500 & -6.50&2.36\\
      D2 & $15 \times 15$ & 15,637,500 & -6.46&2.36\\
      D3 & $15 \times 15$ & 8,400,000 & -6.48&2.30\\ \hline
	\end{tabular}}
\label{table:verification} 
\end{center}
\end{table}

\section{Uncontrolled flow}
\label{sec:Baseline}

\begin{figure*}
\centering
\includegraphics[scale=1.2]{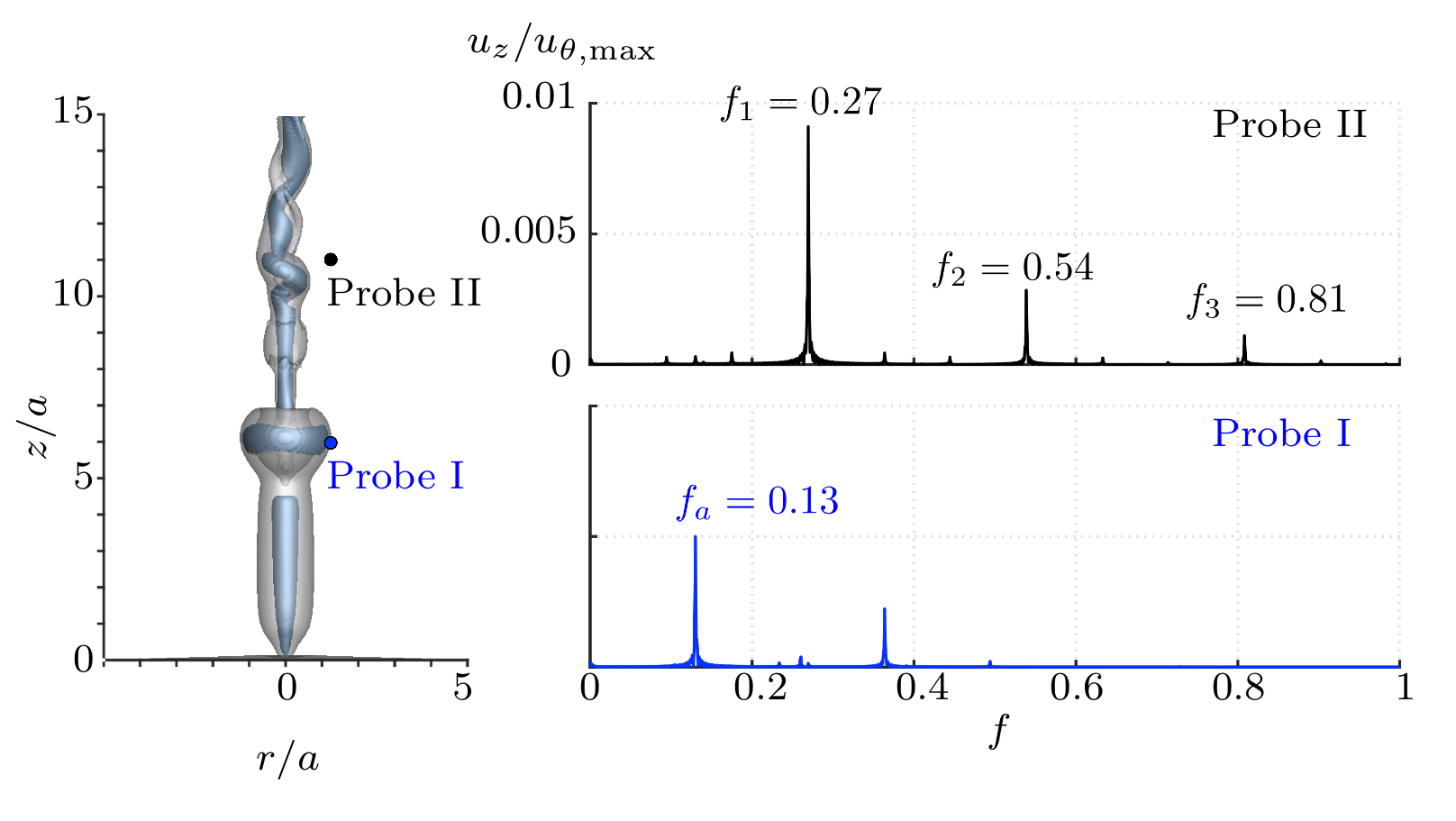}
\caption{ (\emph{Left}) Instantaneous vortical flow visualized with $Q$-criterion of $Q=2$ (blue)  and vorticity magnitude of $||\omega||=2$ (grey).  (\emph{Right}) Spectra for ($u_z/u_{\theta, \max}$) for Probe I at $(r/a, z/a)=(1, 6)$ and Probe II at (1, 11).}
\label{fig:Q_criterion_base} 
\end{figure*}

Let us first present the uncontrolled base flow and characterize its dynamical properties.  The vortical flow that is considered here presents distinct flow structures, as visualized by the $Q$-criterion \citep{hunt1988eddies} isosurface in Fig.~\ref{fig:Q_criterion_base}.  By examining the frequency content of the flow, we classify the vortical structures into three regions: (1) \emph{steady vortex}, (2) \emph{vortex breakdown} and (3) \emph{wake} regions.
The steady vortex emerges at the lower axial location of $0<z/a<5$. Its vortical structure exhibits an axisymmetric profile. 
Over the region of $5<z/a<7$, the vortical structure bulges and forms a bubble-like vortex breakdown.
Downstream of the vortex breakdown ($z/a>8$), the vortex loses its axisymmetric structure and forms an unsteady swirling wake pattern. 
The spectra of the velocity probes identify the oscillation frequencies arising from different regions, which are used for the identification of the three regions. Four dominant frequencies are identified in the vortex breakdown and wake regions. This can be seen in Fig.~\ref{fig:Q_criterion_base}, which shows frequency spectra for two representative positions.  Due to the time-invariant nature of the flow, we do not discuss the lower steady region.  In the vortex breakdown region, we detect a low magnitude oscillation frequency of $f_a=0.13$ and its harmonic frequency of $f=0.36$.  On the other hand, in the wake region, the frequency $f_1=0.27$ and its high-order harmonics $f_2=0.54$ and $f_3=0.81$ are identified.

%%% FIGURE 3 %%%

\begin{figure*}
\centering
\includegraphics[width=\textwidth]{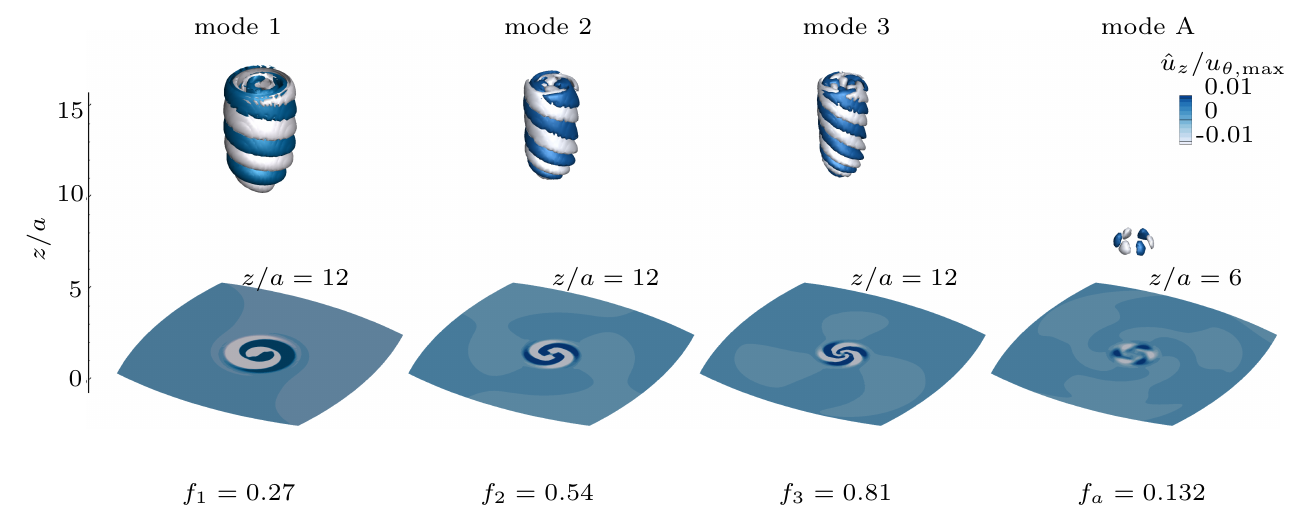}
\caption{Real axial velocity component of DMD modes corresponding to the dominant frequencies in the flow. Projected on the bottom walls are the $r-\theta$ planes at indicated axial slices.}
\label{fig:spectral_baseline} 
\end{figure*}

To further examine the base flow, we perform dynamic mode decomposition (DMD) \citep{schmid2010dynamic,rowley2009spectral,kutz_textbook,taira2017modal} to identify the coherent structures associated with the detected frequencies. 
To conduct DMD of the three-dimensional flow field, we collect 250 snapshots over eight periods for the frequency of $f_1=0.27$. The DMD analysis reveals characteristic structures associated with frequencies $f_1$, $f_2$, $f_3$, and $f_a$, as shown in Fig.~\ref{fig:spectral_baseline}.  Shown are the structures visualized by the real axial velocity component associated with the detected frequencies. 
In the wake region, DMD modes extract helical structures with the dominant frequency $f_1=0.27$ corresponding to the azimuthal wavenumber of $m=1$. 
For the higher frequency DMD modes at $f_2 = 0.54$ and $f_3 = 0.81$, we observe $m=2$ and $3$ helical modes, respectively.
On the other hand, for the vortex breakdown region with $f_a=0.132$, we obtain the non-helical structure, which we refer to as mode A.

%%% FIGURE 4 %%%

\begin{figure*}
\centering
\includegraphics[width=0.65\textwidth]{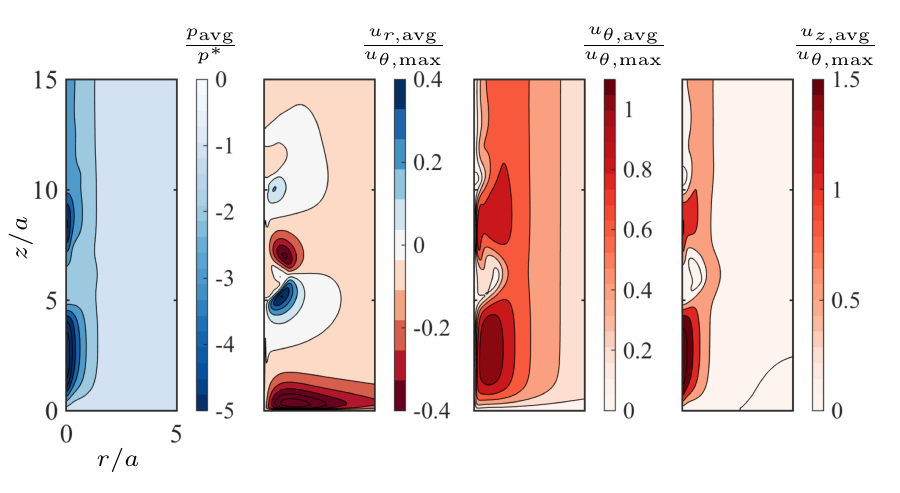}
\caption{Time and azimuthal averaged baseline flow fields}
\label{fig:flowfield_baseline_avg} 
\end{figure*}

The time-averaged flows are shown in Fig.~\ref{fig:flowfield_baseline_avg}.
We observe that as the flow approaches the vortex core, the radially inward momentum is redirected to the axial direction. As a consequence, the vortex is stretched into the axial direction and intensifies its strength.  
The lowest pressure in the time-averaged flow field appears in the steady vortex region, which pertains to the strong swirl motion for this regime.  
This swirling motion is supported by the large inward radial flow that forms along the wall boundary. 
As the flow convects downstream (upward), the emergence of negative azimuthal vorticity leads to the breakdown of the vortex over the region of $5<z/a<7$. The radial velocity shifts its direction at the outer layer of the bubble-like breakdown structure. This dramatic change in the flow triggers unsteadiness into the wall-normal vortical flow as it convects downstream. This velocity distortion weakens the vortex strength, and the core pressure increases in the wake region ($z/a>7$).  
By examining the distribution of the time-averaged pressure, we observe that the core pressure in the wake region is higher than that of the steady vortex region. 

%%% FIGURE 5 %%%

%%%%%%%%%%%%%%%%%%%%%%%%%%%%%%%%%%%%%Active flow control setup%%%%%%%%%%%%%%%%%%%%%
\section{Active flow control setup}
\label{sec:ctrl_setup}
Based on the detailed analysis of the baseline flow from the previous section, we observe two key features that are important in influencing the pressure distributions of the core region.  The first feature is the unsteadiness of the vortical flow.  With the breakdown and the wake instabilities in the flow, we now observe that the low pressure can be alleviated in a time-averaged manner.  The second feature is the strong radial inflow that forms at the bottom wall and strengthens the vortex core. We consider the modification of the vortical structures by enhancing the unsteadiness in the flow and disrupting the strong profile of the axial inflow into the vortex core.

To modify the above two key features of the vortices, we introduce external forcing to the vortical flow from the bottom wall.  We consider the use of mass blowing based actuation, which is known to be effective for a range of flow control applications \citep{Lachmann1961, Gad-el-Hak2000, Taira:AIAAJ09, Munday:AIAAJ18}.
Triggering unsteadiness to modify the vortex velocity profile for pressure increase has been considered in our previous study \citep{liu2018core}.  While the previous approach was found to be effective in modifying the pressure profile, the control mechanism was based on an unsteady input, which can be a challenge to implement in practical applications.
Here we seek a practical control setup with simple steady blowing from the bottom wall.

In the present numerical simulations, we introduce actuation through a boundary condition at the bottom wall.  The boundary condition for actuation is prescribed with a wall-normal velocity profile of 
\begin{equation}
    u^c_z = A\exp{(-r^2/a^2)},
    \label{eq:steady_act}
\end{equation}
where the amplitude of the control actuation is $A$. 
We also consider the robustness of the actuation setup in terms of the actuator location with respect to the core position.  This will be studied by shifting the position of actuation away from the center of the vortex.  
   
We quantify the control effort with the momentum coefficient defined as
\begin{equation}
C_\mu = \frac{\int_0^{2a} 2\pi r {u_z^c}^2 {\rm d}r }{\int_0^{2a} 2\pi r u_{z^\ast}^2(r) {\rm d}r}, 
\end{equation}
where $a$ is the radius of the vortex core, $u_{z^\ast}(r)$ is the time and azimuthal averaged axial velocity profile over radial direction at $z/a=2.3$, where the maximum baseline axial velocity is achieved along the vortex core. The axial velocity from actuation is $u^c_z$.
We vary the control amplitude between $0.05\le A_c \le 0.6$, which corresponds to a momentum coefficient in the range of $0.06\% \le C_\mu \le 9.32\%$ to evaluate the influence of the momentum coefficient on the control effectiveness.
To assess the impact of the present control approaches on the vortex-core pressure, we define the normalized increase in the minimal pressure in the flow
\begin{equation}
    \eta_p = 
    \left| \frac{p_{c,\text{min}}-p_{b,\text{min}}}{p_{b,\text{min}}} \right|,
\end{equation}
where $p_{b,\text{min}}$ and $p_{c,\text{min}}$ are the minimal time-averaged pressure of the base and controlled cases, respectively.

%%%%%%%%%%%%%%%%%%%%%%%%%%%%%%%%%%%%%Steady Controlled flow field%%%%%%%%%%%%%%%%%%%%%

\section{Controlled flow}
\label{sec:ctrl_results}
\begin{figure}
\centering
\includegraphics[width=3.34in]{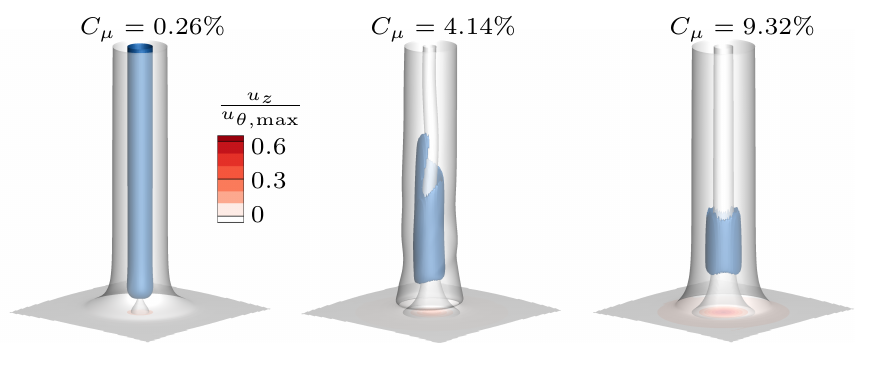}
\caption{Vortical structures modified by steady forcing for $C_\mu=0.26\%, 4.14\%$ and $9.32\%$, visualized with isosurfaces of $||\omega||=0.6$ (grey) and $Q$-criterion of $Q=0.6$ (blue). }
\label{fig:steady_Q} 
\end{figure}

Let us assess the control effect on modifying the wall-normal vortex to increase the pressure distribution within the vortex. The controlled simulation is initialized by the baseline wall-normal vortex. 
We first examine the influence of the amplitude on the wall-normal vortex with $0.06\% \le C_\mu \le 9.32\%$. As shown in Fig.~\ref{fig:steady_Q}, we observe that the steady control input stabilizes the vortical flow.  With the control input placed at the center of the vortex, the vortical flow turns steady with axisymmetric structures along the vortex axis. The steady mass blowing actuation suppresses the vortex breakdown and wake region. As the amplitude increases, the actuation effectively destructs the vortical structure over the steady flow region, as shown in the cases of $C_\mu=4.14\%$ and $9.32\%$.

%%% FIGURE 6 %%%

\begin{figure*}
\centering
\includegraphics[width=\textwidth]{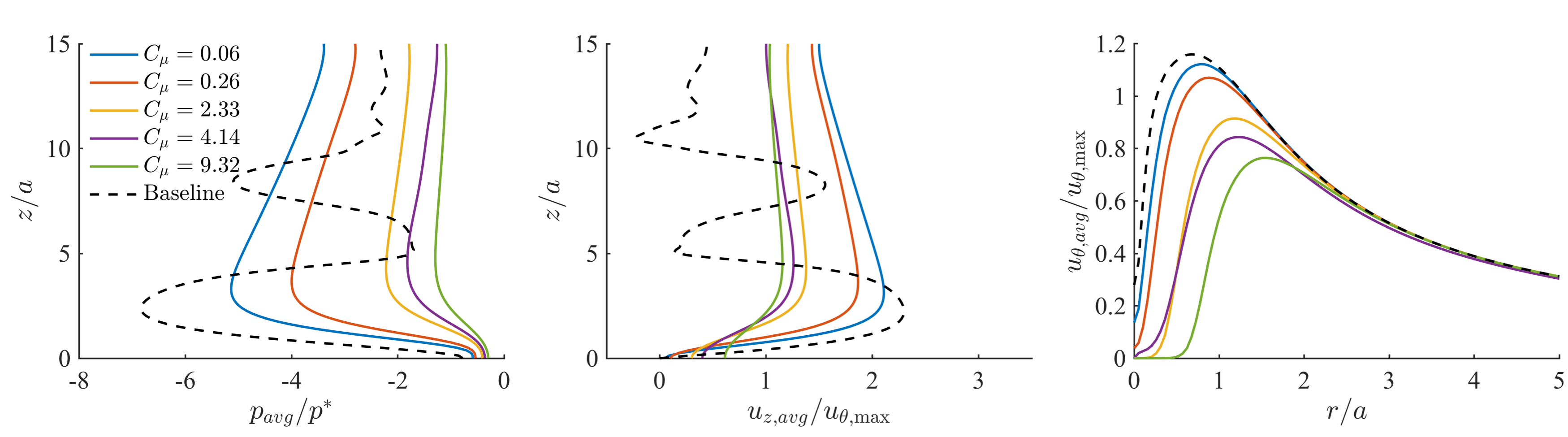}
\caption{Time and azimuthal averaged pressure, axial velocity and azimuthal velocity for the steady controlled cases}
\label{fig:steady_pwv} 
\end{figure*}

Next, we examine the change in the pressure distributions in the vortex for the controlled cases.  As presented in Fig.~\ref{fig:steady_pwv} (left), the pressure distribution along the vortex core changes significantly compared to the uncontrolled flow.
The removal of the vortex breakdown and wake structures makes the pressure evenly distributed along the vortex axis.
With an increase in the strength of blowing, the core pressure increases with its lowest pressure location moving downstream (upward).
The corresponding axial velocity distribution along the vortex axis is shown in Fig.~\ref{fig:steady_pwv} (middle). For all control cases, the double-hump profile of the baseline axial velocity disappears.
We also observe an increase in the magnitude of the axial velocity near the actuator position, as the actuation amplitude increases.
What is noteworthy is that the maximum axial velocity decrease with a smooth profile in the vertical direction as the actuation amplitude increases for $z/a>2$.  This suggests the vertical velocity profile being widened with steady blowing.  This variation due to the increased actuation amplitude reduces the strength of the inward radial velocity, which is directed vertically above the origin.  

%%% FIGURE 7 %%%

We show in Fig.~\ref{fig:steady_pwv} (right) the time-averaged azimuthal velocity distribution along the radial direction at $z/a=3.4$, which is the location for the minimal core pressure for the baseline case.
As the control amplitude increases, the maximum azimuthal velocity moves away from the vortex axis, and its magnitude significantly decreases. For the case of $C_\mu=9.32\%$, we observe that the swirl velocity approach zero for $r/a < 0.5$. The actuation intervenes with the swirling flow around the vortex core. 
The control alters the azimuthal flow until $r/a\approx 3$ beyond which the flow approaches the baseline flow profile.
This suggests that the control effect, while locally confined, can be useful in modifying the core pressure profile.

\begin{figure}
\centering
\includegraphics[scale=1]{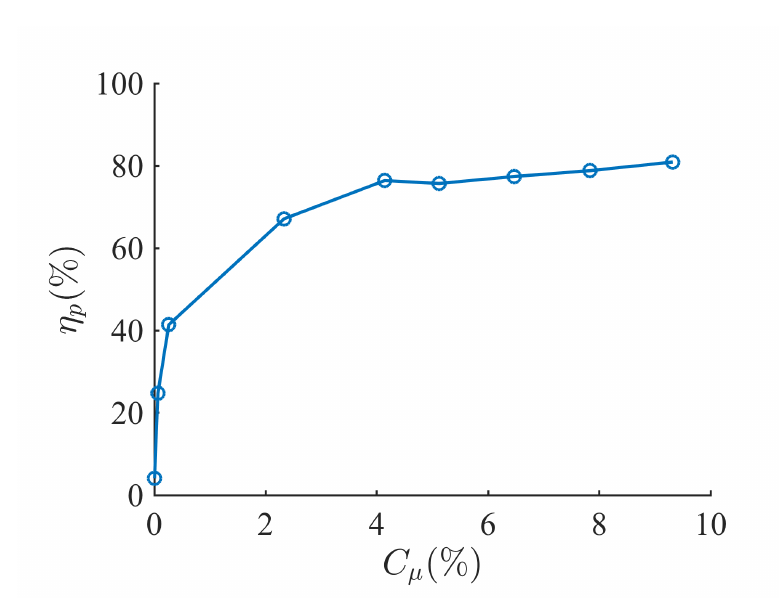}
\caption{Pressure increase over coefficient of momentum}
\label{fig:effect_steady_control} 
\end{figure}

The level of pressure increase achieved by steady blowing is summarized in Fig.~\ref{fig:effect_steady_control}.  When $C_\mu<4.14\%$, the control effect is significantly enhanced as the control input increases. For the case of $C_\mu=4.14\%$, the minimal pressure increases $76.4\%$ compared to the baseline.  For control input past $C_\mu>4.14\%$, control effects are saturated. For this reason, the control setup would not be energetically efficient past that value of forcing.

%%% FIGURE 8 %%%

The present control approach thus far assumes blowing to be introduced at the center of the core.  However, in practical flow settings within a pump sump, the pump-induced vortex can be in motion and off-centered from the pump.  In such cases, it is important that the current control setup can still effectively modify the vortical flow to increase core pressure. For this reason, we examine the effectiveness of steady blowing introduced off-centered from the vortex.  Here, we consider the offset from the vortex center to be $x^*/a=1$, $2$, $3$ and $4$ and consider steady forcing with $C_\mu=4.14\%$. 

\begin{figure}
\centering
\includegraphics[width=3.34in]{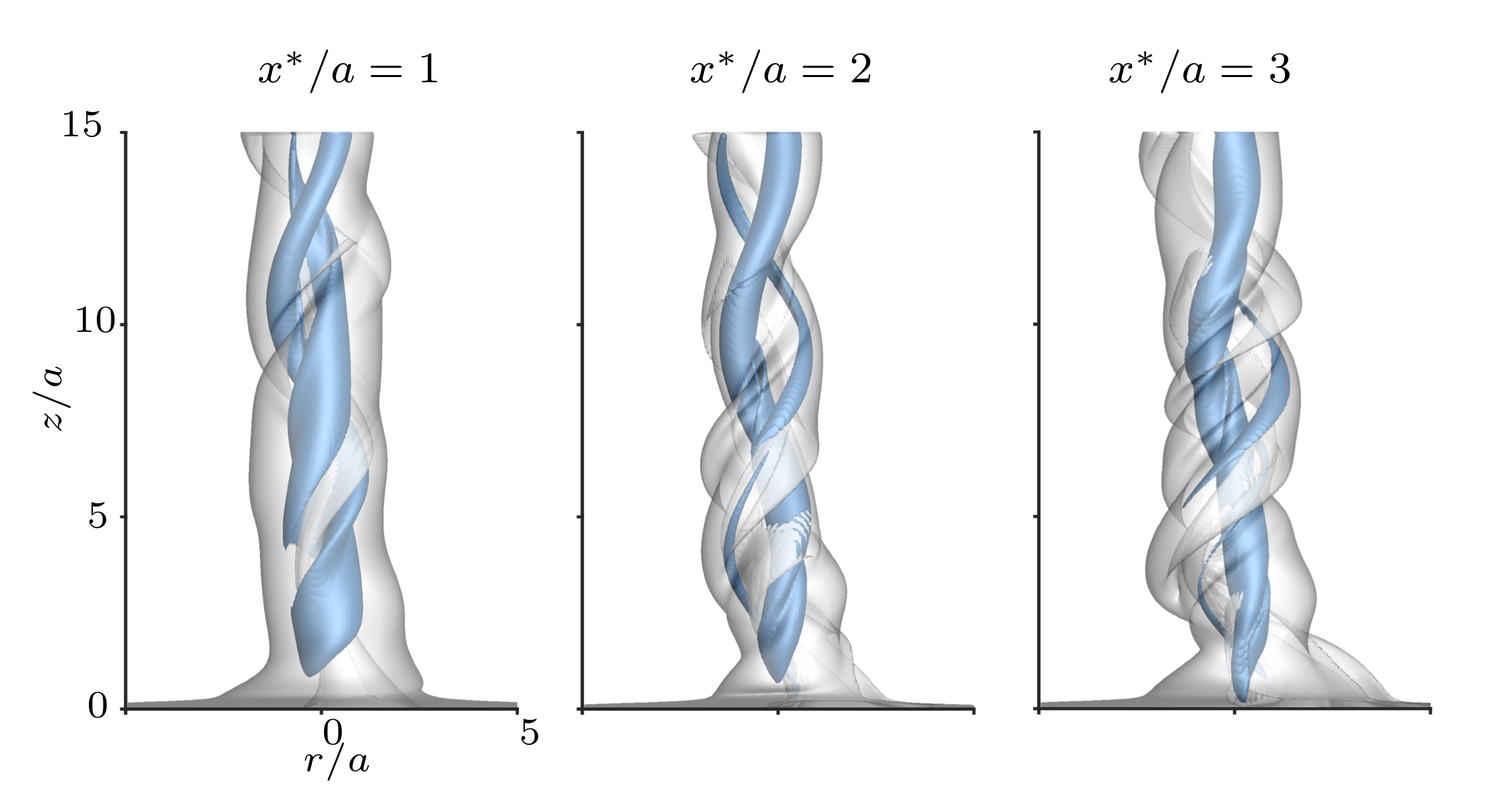}
\caption{Off-centered controlled vortical structures visualized with $Q$-criterion of vorticity magnitude of $||\omega||=0.6$ (gray) and $Q$=0.6 (blue) for $x^*/a=1$, $2$ and $3$}
\label{fig:steady_off_Qcriterion} 
\end{figure}

Let us compare the control effects for the off-centered cases with the centered control case. The off-centered actuation distorts the vortex and forms a steady helical vortex, as shown in Fig.~\ref{fig:steady_off_Qcriterion}.
For all off-centered controlled cases, the flow field exhibits steady helical structures. 
As the actuation moves away from the vortex center, the twisting effect on the primary vortex becomes weak. 
For the case of $x^*/a=3$, the actuation has an impact mostly on the outer layer of the vortex with spiraling structures around the main vortex. 
The modification of the surrounding structure around the vortex weakens the swirl motion of the vortex, and in turn, increases the vortex core pressure. 

\begin{figure}
\centering
\includegraphics[scale=1]{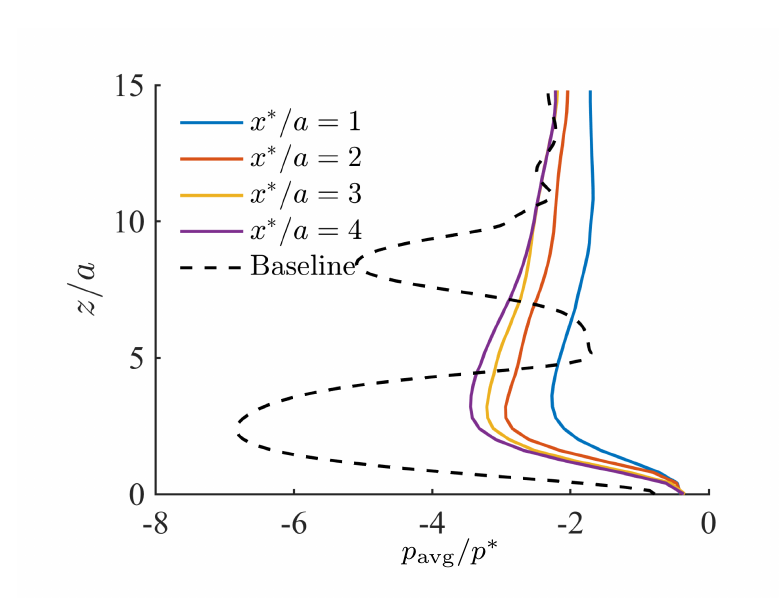}
\caption{Pressure distributions for off-centered control cases}
\label{fig:offs_pressure} 
\end{figure}

The vortex core pressure along the axial direction is presented in Fig.~\ref{fig:offs_pressure}.  As the injection location of steady blowing moves away from the vortex center, the achieved pressure increase sees a reduced benefit. For the controlled cases of $x^*/a=3$ and $4$, the vortex core pressures reach the baseline value downstream of $z/a>10$. In terms of the non-dimensional pressure increase $\eta_p$, the control effect reduces from $66.7\%$ to $49.4\%$ as the actuation moves from $x^*/a=1$ to $4$. 
The results indicate that the present control setup is effective even when the actuation is four core length away from the vortex center.

%%% FIGURE 9 %%%

%%% FIGURE 10 %%%

\begin{figure*}
\centering
\includegraphics[scale=1]{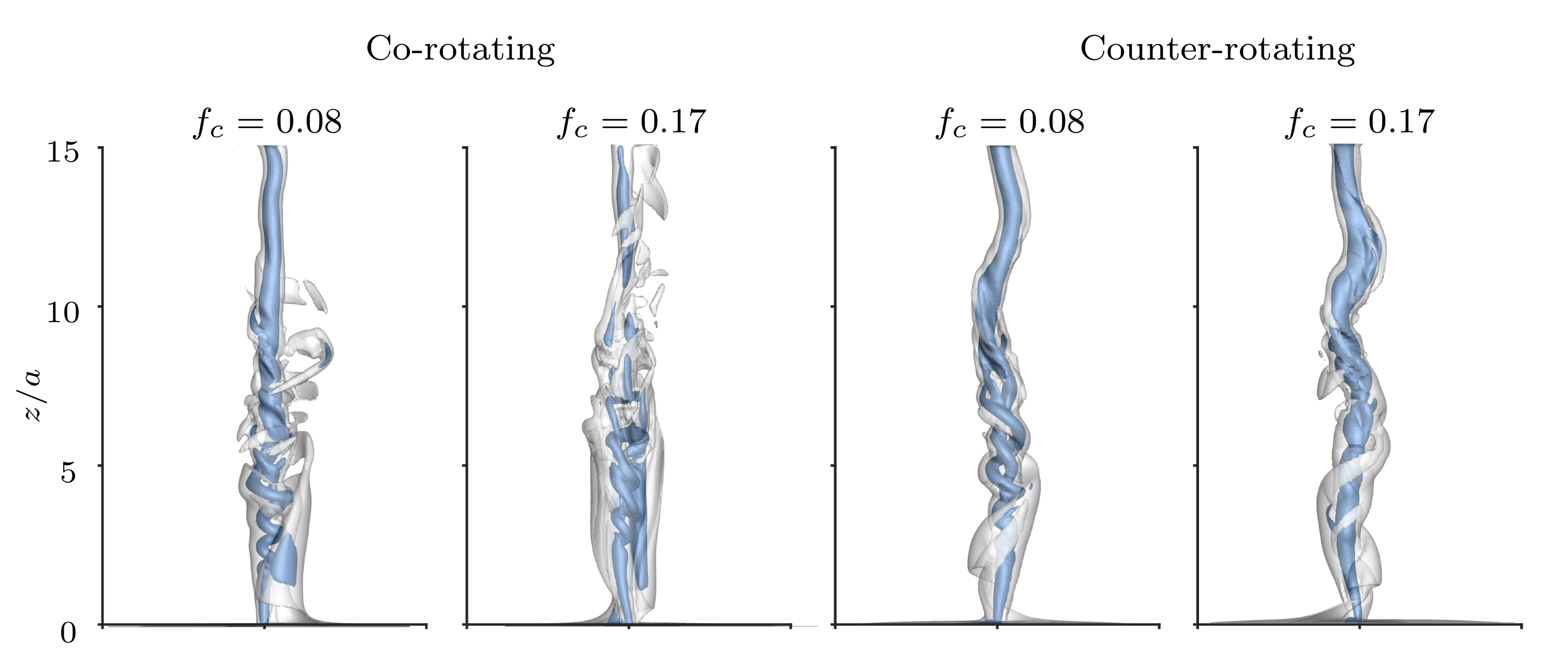}
\caption{Instantaneous vortical structures for unsteady forcing cases with control frequencies of $f_c=0.08$ and 0.17. Iso-surfaces are $Q=2$ (Purple) and $||\omega||=2$ (gray).}
\label{fig:unsteady_Q} 
\end{figure*}

The present results show that the steady mass injection is a robust and effective flow control technique to alleviate the lower core pressure effects. Compared to the previous unsteady rotary flow control\citep{liu2018core}, the underlying control mechanisms are different. 
Fig.~\ref{fig:unsteady_Q} shows the representative vortical structures from co-rotating and counter-rotating unsteady control. 
Here, we define the unsteady momentum coefficient as $$C'_\mu=\frac{f_c \int_0^{1/f_c} \int_0^{2a} \int_0^{2\pi} {u^c_z}^2 r{\rm d}\theta{\rm d}r {\rm d}t}{\int_0^{2a} 2\pi r u^2_{z^\ast}(r) {\rm d}r},$$ where boundary actuation profile\citep{liu2018core} takes the form of $u^c_z=A\cos(2\pi f_c t+m\theta )\exp{(-r^2/a^2)}$ with azimuthal wavenumber $m$ and controlled frequency $f_c$.  
The unsteady actuation introduces instabilities into the vortex. Different types of energy amplification properties, according to the axial wavenumber, affect changing core pressure. 
On the other hand, the steady actuation modifies the inward radial velocity profile close to the vortex core, which lowers the magnitude of the axial velocity and azimuthal velocity to reduce the vortex strength. For this reason, steady control can achieve a significant control effect by increasing the controlled input.

%%% FIGURE 11 %%% 

\begin{figure}
\centering
\includegraphics[width=0.45\textwidth]{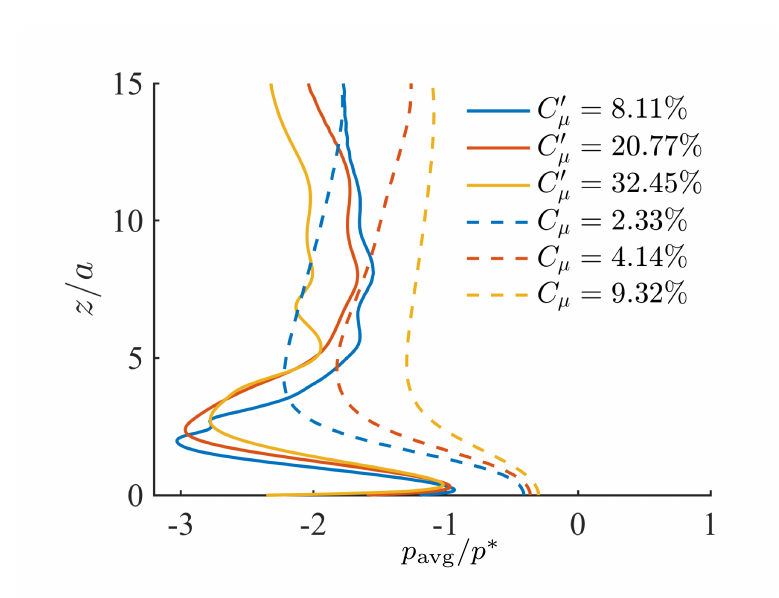}
\caption{The comparison of the vortex core pressure of unsteady (solid lines) and steady (dashed lines) controlled flows.}
\label{fig:unsteady_steady} 
\end{figure}

We show the core pressure distributions along the axial direction in Fig.~\ref{fig:unsteady_steady} for steady control and co-rotating control at $f_c=0.08$, which achieves the best control results by unsteady actuation. For the unsteady actuation, the core pressure distributions indicate that the controlled effect saturates at $C'_\mu=8.11\%$. Indeed, as the control input increases to $C'_\mu=20.7\%$ and $30.4\%$, the core pressure only slightly enhances for $z/a<5$ but decreases for $z/a>5$. 
The steady controlled vortical flow can achieve significant pressure increase as the control input increases. 
The achieved pressure increase for the steady actuation is more significant than what can be attained with unsteady actuation, which suggests the promising performance of steady control actuation. 

%%% FIGURE 12 %%%

\section{Conclusions}
\label{sec:conclusion}

We numerically examined a pump-induced wall-normal single-phase vortex model and applied active flow control for the purpose of increasing the vortex core pressure profile. We found that the steady wall-normal momentum injection is an effective and feasible control technique to achieve the alleviation of the low-pressure core by spreading the core profile. For the case of the momentum coefficient being $C_\mu=2.33\%$ and $4.14\%$, the minimal pressure increases $67.1\%$ and $76.4\%$, respectively, compared to the baseline pressure value. For control input $C_\mu$ above $4.14\%$, the pressure increase saturates. 
The performance of the control setup was also assessed for off-centered cases. We examined the control effect by moving the actuation away from the vortex core. While the effectiveness of the control schemes reduces as the forcing is introduced further away from the vortex center, steady actuation still achieves $49.4\%$ of pressure increase compared to the baseline even when the actuation is four core lengths away from the vortex center. Based on the observations from this study, we find that the simplicity of the steady axisymmetric blowing profile is a promising approach to modify submerged vortices in pump sumps.

\section*{Acknowledgement}

QL and KT thank the support from Ebara Corporation.

\end{document}